\documentclass[aps, prl, superscriptaddress, twocolumn]{revtex4-1}
\usepackage{amsmath,amssymb,amsthm,paralist}
\usepackage{graphicx}
\usepackage{appendix}
\usepackage{subfigure}
\usepackage{longtable}
\usepackage{url}

\newcommand{\isotope}[2]{\ensuremath{\mathrm{^{#1}#2}}}

\begin{document}

\title{Time-of-flight mass measurements for nuclear processes in neutron star crusts}
 
\date{\today}
\author{A. Estrad\'e} 
\altaffiliation{Present address: Saint Mary's University, Canada, and GSI, Germany.}
\affiliation{National Superconducting Cyclotron Laboratory, Michigan State University, USA}
\affiliation{Department of Physics and Astronomy, Michigan State University, USA}
\affiliation{Joint Institute for Nuclear Astrophysics, USA}
\author{M.~Mato\v{s}} 
\affiliation{National Superconducting Cyclotron Laboratory, Michigan State University, USA}
\affiliation{Joint Institute for Nuclear Astrophysics, USA}
\affiliation{Department of Physics and Astronomy, Louisiana State University, USA.}
\author{H.~Schatz} 
\affiliation{National Superconducting Cyclotron Laboratory, Michigan State University, USA}
\affiliation{Department of Physics and Astronomy, Michigan State University, USA}
\affiliation{Joint Institute for Nuclear Astrophysics, USA}

        \author{A.~M.~Amthor} 
\affiliation{National Superconducting Cyclotron Laboratory, Michigan State University, USA}
\affiliation{Department of Physics and Astronomy, Michigan State University, USA}
\affiliation{Joint Institute for Nuclear Astrophysics, USA}
        \author{D.~Bazin} 
\affiliation{National Superconducting Cyclotron Laboratory, Michigan State University, USA}
        \author{M.~Beard} 
\affiliation{Department of Physics, University of Notre Dame, USA}
\affiliation{Joint Institute for Nuclear Astrophysics, USA}
        \author{A.~Becerril} 
\affiliation{National Superconducting Cyclotron Laboratory, Michigan State University, USA}
\affiliation{Department of Physics and Astronomy, Michigan State University, USA}
\affiliation{Joint Institute for Nuclear Astrophysics, USA}
        \author{E.~F.~Brown} 
\affiliation{National Superconducting Cyclotron Laboratory, Michigan State University, USA}
\affiliation{Department of Physics and Astronomy, Michigan State University, USA}
\affiliation{Joint Institute for Nuclear Astrophysics, USA}
	\author{R.~Cyburt} 
\affiliation{National Superconducting Cyclotron Laboratory, Michigan State University, USA}
\affiliation{Joint Institute for Nuclear Astrophysics, USA}
        \author{T.~Elliot} 
\affiliation{National Superconducting Cyclotron Laboratory, Michigan State University, USA}
\affiliation{Department of Physics and Astronomy, Michigan State University, USA}
\affiliation{Joint Institute for Nuclear Astrophysics, USA}
        \author{A.~Gade} 
\affiliation{National Superconducting Cyclotron Laboratory, Michigan State University, USA}
\affiliation{Department of Physics and Astronomy, Michigan State University, USA}
        \author{D.~Galaviz} 
\affiliation{National Superconducting Cyclotron Laboratory, Michigan State University, USA}
\affiliation{Joint Institute for Nuclear Astrophysics, USA}
        \author{S.~George} 
\affiliation{National Superconducting Cyclotron Laboratory, Michigan State University, USA}
\affiliation{Joint Institute for Nuclear Astrophysics, USA}
        \author{S.~S.~Gupta} 
\affiliation{Indian Institute of Technology - Ropar, India}
\affiliation{Joint Institute for Nuclear Astrophysics, USA}
        \author{W.~R.~Hix} 
\affiliation{Physics Division, Oak Ridge National Laboratory, USA}
        \author{R.~Lau} 
\affiliation{National Superconducting Cyclotron Laboratory, Michigan State University, USA}
\affiliation{Department of Physics and Astronomy, Michigan State University, USA}
\affiliation{Joint Institute for Nuclear Astrophysics, USA}  
        \author{G.~Lorusso} 
\affiliation{National Superconducting Cyclotron Laboratory, Michigan State University, USA}
\affiliation{Department of Physics and Astronomy, Michigan State University, USA}
\affiliation{Joint Institute for Nuclear Astrophysics, USA}
        \author{P.~M\"oller} 
\affiliation{Los Alamos National Laboratory, USA}
        \author{J.~Pereira} 
\affiliation{National Superconducting Cyclotron Laboratory, Michigan State University, USA}
\affiliation{Joint Institute for Nuclear Astrophysics, USA}
        \author{M.~Portillo} 
\affiliation{National Superconducting Cyclotron Laboratory, Michigan State University, USA}
        \author{A.~M.~Rogers} 
\affiliation{National Superconducting Cyclotron Laboratory, Michigan State University, USA}
\affiliation{Department of Physics and Astronomy, Michigan State University, USA}
\affiliation{Joint Institute for Nuclear Astrophysics, USA}
        \author{D.~Shapira} 
\affiliation{Physics Division, Oak Ridge National Laboratory, USA}
        \author{E.~Smith} 
\affiliation{The Ohio State University, USA}
\affiliation{Joint Institute for Nuclear Astrophysics, USA}
        \author{A.~Stolz} 
\affiliation{National Superconducting Cyclotron Laboratory, Michigan State University, USA}
        \author{M.~Wallace} 
\affiliation{Los Alamos National Laboratory, USA}
        \author{M.~Wiescher} 
\affiliation{Department of Physics, University of Notre Dame, USA}
\affiliation{Joint Institute for Nuclear Astrophysics, USA}

\begin{abstract}
The location of electron capture heat sources in the crust of accreting neutron stars depends on the masses of extremely neutron-rich nuclei. We present first results from a new implementation of the time-of-flight technique to measure nuclear masses of rare isotopes at the National Superconducting Cyclotron Laboratory. The masses of 16 neutron-rich nuclei in the scandium -- nickel range were determined simultaneously,  improving the accuracy compared to previous data in 12 cases. The masses of  $^{61}${V}, $^{63}${Cr}, $^{66}${Mn}, and $^{74}${Ni} were measured for the first time with mass excesses of $-30.510(890)$ MeV, $-35.280(650)$ MeV, $-36.900(790)$ MeV, and $-49.210(990)$ MeV, respectively. With the measurement of the $^{66}$Mn mass, the locations of the two dominant electron capture heat sources in the outer crust of accreting neutron stars that exhibit superbursts are now experimentally constrained. We find that the location of  the $^{66}$Fe$\rightarrow^{66}$Mn electron capture transition occurs significantly closer to the surface than previously assumed because our new experimental Q-value is 2.1~MeV (2.6$\sigma$) smaller than predicted by the FRDM mass model. 
\end{abstract}
 
 \maketitle
 
Neutron stars that accrete matter from an orbiting low-mass companion star are observed as galactic X-ray binaries \cite{Strohmayer_06}. A fluid element accreted onto the neutron star surface is buried by the continuous accretion of more matter, and  undergoes a sequence of compositional transformations driven by nuclear reactions under rising pressure. Near the surface, at typical depths of a few meters, thermonuclear explosions, which are observed as X-ray bursts, burn hydrogen and helium into heavier elements in the nickel -- cadmium range \cite{Schatz_01}. In somewhat deeper layers explosive 
carbon burning is thought to produce the occasionally observed superbursts \cite{Cumming_01}, converting the ashes of the regular bursts into nuclei in the iron -- nickel range. Still deeper in the neutron star crust the matter undergoes a sequence of electron captures, accompanied at even greater depth by neutron emissions, and pycnonuclear 
fusion reactions \cite{Gupta_07,Gupta_09,Haensel_Zdunik_08}. These nuclear processes, which involve extremely neutron-rich nuclei, heat the crust creating a characteristic temperature profile.  

In this letter we report results from an experiment at the National Superconducting Cyclotron Laboratory (NSCL), where we have produced and measured the masses of neutron-rich nuclei using a new implementation of the time-of-flight (TOF) technique. We measured the mass of sixteen neutron-rich isotopes in the region around $N=40$, four of which were measured for the first time.
The results allow us to locate the dominant electron capture heat sources in the crust of accreting neutron stars and better constrain their strength. The masses also provide new information on the onset of deformation near the $N=40$ region in neutron-rich nuclei. 

An understanding of crustal heating nuclear processes in accreting neutron stars  is needed to interpret a number of observables. 
Neutron stars in transiently accreting X-ray binaries offer the unique opportunity to directly observe the temperature profile of the crust. Some systems accrete for many years, sufficiently long to reach thermal equilibrium. Then accretion stops for many years, enabling the observation of the thermally relaxing crust over time (for example Ref. \cite{Cackett_08}). The time dependence 
of the cooling curve contains information on crust properties such as composition, thermal conductivity, heat capacity, neutron superfluidity, and the efficiency of neutrino cooling \cite{Shternin_07,Brown_09}. Interpretation of these observations requires reliable nuclear physics to predict the location and strength of the nuclear heat sources during the accretion phase. Crustal heating predictions are also needed to understand the recurrence time of  superbursts \cite{Cumming_01}, and the generation of gravitational waves due to the deformations induced by electron capture reactions in the crust of the rapidly spinning neutron star, which might be observable with future gravitational wave detectors  \cite{Bildsten_98}.

\begin{figure}[tb]\begin{center}
\includegraphics[width=0.95\columnwidth]{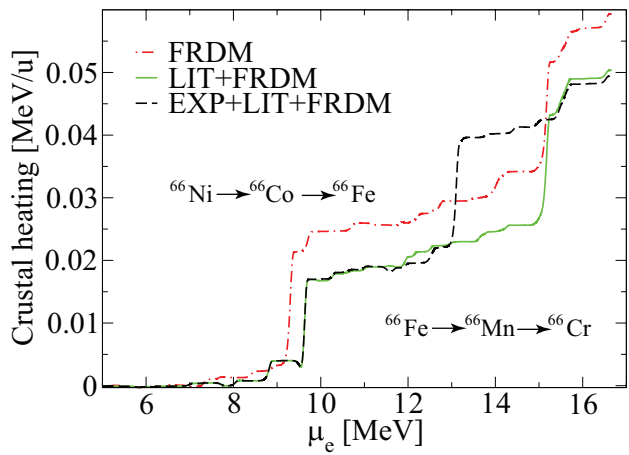}
\caption{
Integral of the heat deposited in the neutron star crust by electron captures as a function of depth (indicated through $\mu_{\rm e}$). Shown are results for masses from FRDM \cite{FRDM} only (red, dot-dashed), for
implementing previously published experimental masses (green, solid), and for implementing in addition our new mass results (dashed, black).} \label{fCrustHeating}  
\end{center}\end{figure}

Which electron captures occur in the outer crust of an accreting neutron star depends on the composition synthesized by thermonuclear burning processes at the neutron star surface. Model calculations show that for systems exhibiting superbursts, resulting ashes are mainly nuclei in the $A=54$ -- 66 mass range \cite{Schatz_03}. These ashes serve as the initial composition for the electron capture processes. Crust model calculations \cite{Gupta_07}  show that for such composition, heat release in the outer crust is dominated by two transitions: two-step electron captures on \isotope{66}{Ni} $\rightarrow$ \isotope{66}{Co} $\rightarrow$ \isotope{66}{Fe}, and subsequently on \isotope{66}{Fe} $\rightarrow$ \isotope{66}{Mn} $\rightarrow$ \isotope{66}{Cr}  (corresponding to steep increases in depth-integrated heat in Fig.~\ref{fCrustHeating}). The location of these transitions is set by their electron capture threshold  $Q_{\rm EC}$. Because the electron chemical potential $\mu_{\rm e}$ slowly rises with depth and the temperature is rather low ($kT \lesssim 40$~keV$\ll  \mu_{\rm e}$) the transition occurs essentially at a depth where $\mu_{\rm e} \approx  Q_{\rm EC}$.  $Q_{\rm EC}=\Delta(Z,N)-\Delta(Z-1,N+1)-E_{\rm x}$ depends on the mass excess $\Delta$ of parent and daughter nuclei, and the excitation energy of the lowest lying state $E_{\rm x}$ into which the capture can occur. Because of nuclear pairing there is a strong odd-even staggering of  $Q_{\rm EC}$ and the threshold for the two-step transition is effectively set by the first step, the electron capture on \isotope{66}{Ni} and \isotope{66}{Fe}. In both cases $E_{\rm x}$ is predicted to be negligible (of the order of 0.1~MeV), and therefore it is the nuclear masses of $^{66}$Ni, $^{66}$Co, $^{66}$Fe, and $^{66}$Mn that determine where the heat is deposited. With our first mass measurement of  $^{66}$Mn, all these masses are now known experimentally.

TOF mass measurements have been successfully applied with different technical approaches at several facilities for the study of short-lived isotopes \cite{Savajols_01, Lunney_03}. We report here a first implementation of the TOF technique at the NSCL. Details of the setup and the analysis will be presented in an upcoming publication. Neutron-rich isotopes were produced by fragmentation of a $^{86}$Kr primary beam at 100 MeV/u in a Be target. The fragments were collected by the A1900 fragment separator and transmitted to the S800 spectrometer \cite{Bazin_03} through a beamline. Two production targets, with thicknesses of 51~mg/cm$^2$ and 94~mg/cm$^2$, were alternated, keeping the magnetic rigidity of the A1900, the beamline, and the S800 unchanged, to increase the transmission of particles with different mass-to-charge ratios. Thus we obtain a sufficient number of calibration nuclei with well-known masses. The TOF was measured with newly developed fast timing scintillators located at the focal planes of the A1900 and S800 resulting in a flight path of 58.7~m. The momentum acceptance of the system was 0.5 \%, requiring a precise relative magnetic rigidity ($B \rho$) measurement of each beam particle. This was accomplished with a position sensitive micro-channel plate detector located at a dispersive focus of the S800. Detectors at the S800 focal plane provided energy loss measurements for particle identification and beam tracking information.

From the simultaneous measurement of magnetic rigidity, TOF, and atomic charge number (from energy loss) for each fully-stripped beam ion, the mass can be determined. The measured TOF of each isotope was corrected for its dependence on the measured $B\rho$ using an empirical relationship. The resulting relative mass resolution was $1.8 \times 10^{-4}$ for the typical case.

\begin{figure}[tb]\begin{center}
\includegraphics[height=0.95\columnwidth, angle=-90]{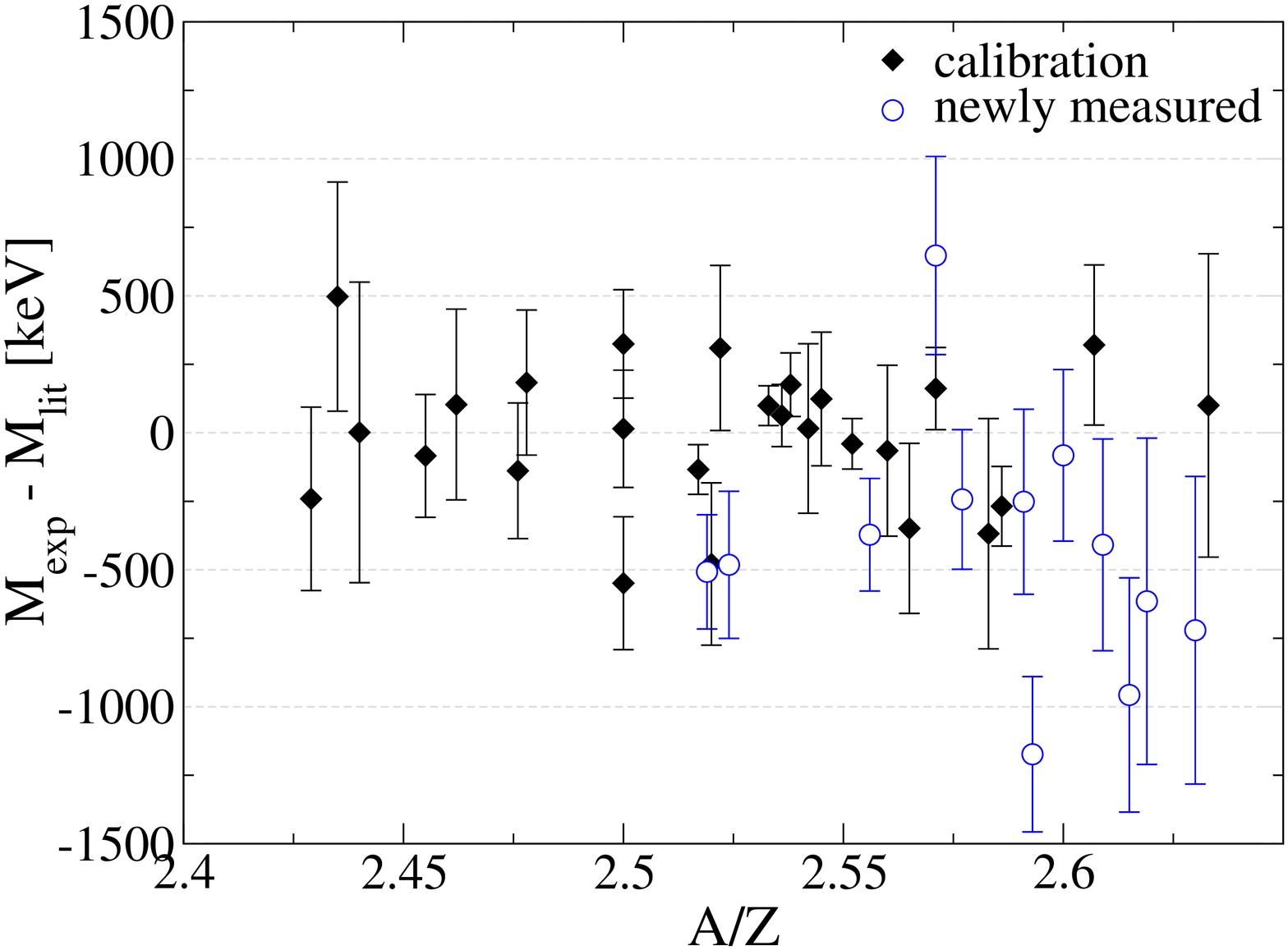}
\caption{Black diamonds show fit residuals for the reference isotopes as a function of their mass-to-charge ratio ($^{51}$Sc, $^{52}$Sc, $^{54}$Ti, $^{55}$Ti, $^{56}$Ti, $^{56}$V, $^{57}$V, $^{58}$V, $^{59}$V, $^{60}$Cr, $^{61}$Cr, $^{62}$Cr, $^{61}$Mn, $^{63}$Mn, $^{64}$Mn, $^{64}$Fe, $^{66}$Fe, $^{71}$Ni, $^{72}$Ni, $^{73}$Ni, $^{73}$Cu, $^{74}$Cu, $^{75}$Cu, $^{75}$Zn, $^{76}$Zn, and $^{79}$Zn). Blue circles are the isotopes for which we present an improved mass value (see Table \ref{tResults}).}
\label{ffitFINAL}
\end{center}\end{figure}

The relation between TOF and $m/q$ of each ion was obtained by fitting a 6 parameter calibration function of second order in TOF and third order in $Z$ to the measured TOFs of 26 reference isotopes of known mass \cite{AME03, Guenaut_07,Rahaman_07,Ferrer_10}.  The reference masses included two isotopes with known low-lying isomers ($^{64}$Mn and $^{75}$Cu \cite{Daugas_10}).  We have confirmed that the unknown population of these isomers does not affect the final results, by performing different fits with variations of the $^{64}$Mn and $^{75}$Cu masses that account for a range of possible isomeric populations, following the method in Ref.~\cite{AME03}.  The resulting fit residuals show no apparent systematic trends (Fig.~\ref{ffitFINAL}) but the $\chi^2$ per degree of freedom of the fit is larger than one, indicating the presence of additional systematic errors. To estimate the magnitude of these errors we find the additional error that normalizes $\chi^2$ per degree of freedom to one when added in quadrature to each m/q calibration data point. We find a systematic error of 5.3 keV$/q$, or 130 keV for manganese. To determine the error of a measured mass, we add in quadrature this systematic error, the statistical error, and the calibration error from the uncertainties of the fit parameters explicitly calculated from the fit covariance matrix. Results are shown in Table~\ref{tResults}. For the new masses in this work the statistical error dominates, with the calibration error contributing a significant fraction.

\begin{table}[tb]
\centering
\caption{Mass excess results, in keV, from the present experiment and from the literature \cite{AME03} (all corresponding to measurements at TOFI \cite{TOFI}). The fourth column shows the counts of each isotope used for our measurement.}
\medskip
\begin{tabular}{cccr}
\hline
\hline
 & This work & Literature & N(events)\\
\hline

\isotope{53}{Sc}& -38110 (270)&	-37630 (280)  &  6000 \\ 
\isotope{54}{Sc} \footnote{Isotopes with known long-lived isomers with energies that range up to 440 keV \cite{NUBASE03}.}& -33540 (360) &	-34190 (370) & 1700 \\ 	
\isotope{55}{Sc}&	-30240 (600) &	-29620 (750) & 500 \\
\isotope{57}{Ti}&	-33790	(340) &	-33530 (470) & 1700 \\ 
\isotope{60}{V} \footnotemark[1]&	-33010	(390) &	-32600 (470)  & 1500 \\
\isotope{61}{V}&	-30510	(890)& -- &  300 \\
\isotope{63}{Cr}&	-35280	(650) & -- & 600 \\
\isotope{65}{Mn}&	-40790	(310) &	-40710 (560) & 3100 \\
\isotope{66}{Mn}&	-36900	(790) & -- & 400 \\ 
\isotope{67}{Fe} \footnotemark[1]&	-45980	(250) &	-45740 (370) & 5300 \\
\isotope{68}{Fe}&	-44090	(430) &	-43130 (750) & 1500 \\ 
\isotope{68}{Co} \footnotemark[1]&	-51860	(210) &	-51350	(320) & 17200 \\
\isotope{69}{Co}&	-50370	(210) &	 -50000	(340) & 15500 \\
\isotope{70}{Co} \footnotemark[1]&	-46820	(280) &	-45640 (840) & 4900 \\
\isotope{71}{Co}&	-44590	(560) &	-43870 (840) & 1100 \\
\isotope{74}{Ni}&	-49210	(990) & -- 	& 300 \\ 
\hline
\hline
\end{tabular}
\label{tResults}
\end{table}

\begin{figure}[tb]\begin{center}
\includegraphics[height=0.95\columnwidth, angle=-90]{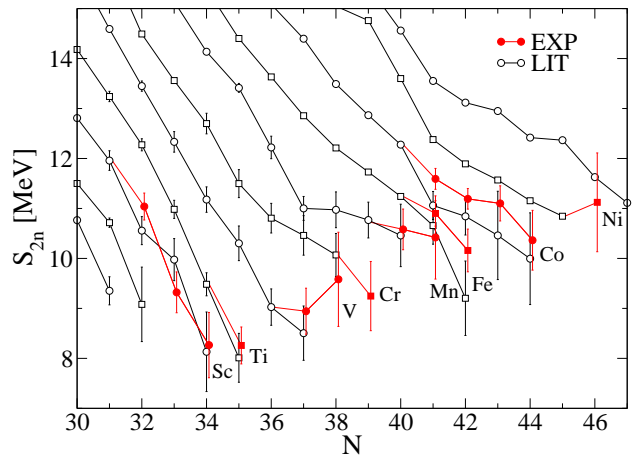}
\caption{Two neutron separation energies as a function of neutron number. $S_{\rm 2n}$ values calculated with our masses results are shown as filled symbols (red). Open symbols show $S_{\rm 2n}$ values from the literature \cite{AME03,Guenaut_07,Rahaman_07,Ferrer_10}.}
\label{fS2n}
\end{center}\end{figure}

Figure~\ref{fS2n} shows the systematics of two-neutron separation energies ($S_{\rm 2n}$) extended towards more neutron-rich nuclei by our measurements. As the fp-shell gets filled with neutrons,  configurations involving the $\mathrm{g}_{9/2}$ shell start driving increased deformation near $N=40$ for $Z<28$ nuclei leading to increased binding energies, and a change in the slope of the observed two-neutron separation energies as functions of neutron number  \cite{Lenzi_10}. For iron, $\gamma$-spectroscopy studies have demonstrated a marked decrease in the energy of the first 2$^+$ state in even-even isotopes starting at $N=38$ indicating the onset of increased deformation, coinciding with the slope change in $S_{\rm 2n}$ \cite{Hannawald_99, Adrich_08}. A similar effect is observed for chromium starting at $N=36$, again confirmed by $\gamma$-spectroscopy studies \cite{Sorlin_03, Gade_10}. For manganese and iron our new masses confirm the continuation of this trend beyond $N=40$. Interestingly, for vanadium isotopes our new mass of $^{61}$V shows the onset of the same effect beyond $N=36$. While our uncertainty for $^{61}$V is large (890~keV) the deviation from the linear trend of the  lighter vanadium isotopes is about 3.5~MeV, or 	4$\sigma$. This is in line with a comparison of experimental  $\beta$-decay half-lives with shell model calculations, which indicated that for $^{61}$V a pure fp-shell model calculation is not adequate, hinting at the onset of the influence of the $\mathrm{g}_{9/2}$ shell \cite{Gaudefroy_05}. For the cobalt isotopes, our mass results are systematically lower than previous measurements with the TOFI spectrometer \cite{TOFI}. The same trend was observed in recent Penning trap measurements for $^{66}$Co and $^{67}$Co \cite{Ferrer_10}. 

Using the new masses we carried out model calculations of the heating in the crust of an accreting neutron star, using the same steady state crust model described in detail in \cite{Gupta_07}. The initial composition is the ashes of a carbon superburst occurring at a depth where $\mu_{\rm e} \approx 4$~MeV. Shown in Fig.~\ref{fCrustHeating} are results using the FRDM  mass model \cite{FRDM}, and results when including available experimental masses. The FRDM was employed here, as well as in previous crust model calculations \cite{Gupta_07}, for consistency with the QRPA approach used to calculate electron capture rates. The heating is dominated by two sources: electron capture on $^{66}$Ni (at $\mu_{\rm e}\approx9.4$~MeV with FRDM), and on $^{66}$Fe (at $\mu_{\rm e}\approx15.3$~MeV with FRDM). Previous mass measurements allow one to pinpoint the electron capture on $^{66}$Ni, shifting it slightly deeper to $\mu_{\rm e}\approx9.6$~MeV. The energy this transition releases in the crust is smaller because of a smaller odd-even staggering in the experimental Q-values (for $^{66}$Co the experimental $Q_{\rm EC}=-6.34\pm0.02$~MeV, while the FRDM value is $Q_{\rm EC}= -5.40$~MeV). Our new result of the $^{66}$Mn mass allows us now to also place the $^{66}$Fe transition based on an experimental $Q_{\rm EC}$. Because our new  $Q_{\rm EC}$ for  $^{66}$Fe of $-13.2\pm0.8$~MeV  is
2.1~MeV (2.6$\sigma$) smaller than the FRDM prediction, the $^{66}$Fe transition turns out to occur at a much shallower depth, around $\mu_{\rm e}\approx13.2$~MeV.

A possible reason for the discrepancy of the FRDM prediction of the $Q_{\rm EC}$ for $^{66}$Fe is that the spherical $Z$=28 shell gap is too strong in the model preventing the onset of deformation at $N$=40. We can also compare our new $^{66}$Fe $Q_{\rm EC}$ result with Q-values obtained with the microscopic mass models HFB-14 \cite{HFB14} and HFB-17 \cite{HFB17}.  HFB-14 predicts $Q_{\rm EC}=-12.2$~MeV, more than 1$\sigma$ different from the experimental value. On the other hand, the recent HFB-17 predicts $Q_{\rm EC}=-13.2$~MeV in agreement with experiment, though the individual masses are each discrepant by about 0.9~MeV.  However, the error estimates of the $Q_{\rm EC}$ theoretical predictions (corrected for experimental errors following the procedure in \cite{FRDM}, Eq. (6)) in the entire region of interest (even $A$ chains with $39 < A < 71$ and $Q_{\rm EC} < -3$ MeV) for FRDM, HFB-14, and HFB-17 are 0.76~MeV, 0.89~MeV, and 0.85~MeV, respectively.

In Fig.~\ref{fCrustHeating} our new masses lead to a reduction of the total heat produced compared to \cite{Gupta_07} because the $^{66}$Fe transition occurs at a lower $\mu_{\rm e}$, leading to a smaller $\mu_{\rm e}-Q$ for the capture on $^{66}$Mn that directly follows. The $Q_{\rm EC}$ for $^{66}$Mn is still taken from FRDM, potentially leading to an unrealistic odd-even staggering. A mass measurement of $^{66}$Cr is needed to address this issue. In addition, the amount of crustal heating depends on the prediction of the correct excitation energies for the final states for electron captures on $^{66}$Co and $^{66}$Mn, which affects the fraction of energy lost by neutrino emission. 

Crustal heating in accreting neutron stars can be strongly affected by nuclear structure effects associated with sub-shell closures \cite{Gupta_07}. Our results for $N$=40 show that at the same time, mass models can be particularly uncertain in these regions. A similar effect might be expected for neutron rich nuclei with $A\sim$100 near $N$=60 \cite{RGuzman_10}, which would be relevant for models of heating in neutron star crusts with a subtantial amount of heavier X-ray burst ashes.

In summary, we have presented the first results from a new implementation of the TOF technique at the NSCL to measure masses of very neutron-rich nuclei. Systematic errors of approximately 130~keV have been achieved, and the resolution is sufficient for measurements with a few 100~keV accuracy for nuclei where a few 1000 events can be produced. Our measurement of the $^{61}$V mass indicates increased deformation and the beginning of influence of the $\mathrm{g}_{9/2}$ orbital in this isotopic chain. Our first measurement of the $^{66}$Mn mass allows us to determine the depth of all the major heat sources in the outer crust of accreting neutron stars. We find that one of them, the electron capture on $^{66}$Fe, occurs at a much shallower depth than predicted. The 800~keV uncertainty of the new electron capture Q-value for $^{66}$Fe is much smaller than typical variations in mass model predictions. Furthermore, the uncertainty is now experimentally determined and can be considered in astrophysical models in a quantitative way. With the depth of the transitions fixed, the remaining smaller nuclear physics uncertainties affecting the amount of heat deposited can now be taken into account when using models of crustal heating to interpret observations. A mass measurement of $^{66}$Cr would be helpful to reduce these uncertainties further. 

\begin{acknowledgments} 
This work was partially supported by NSF grants 08-22648 and PHY 06-06007, and by DFG under contract number GE 2183/1-1.
\end{acknowledgments}

\end{document}